# FIRST 3D PRINTED IH-TYPE LINAC STRUCTURE - PROOF-OF-CONCEPT FOR ADDITIVE MANUFACTURING OF LINAC RF CAVITIES

H. Haehnel*, U. Ratzinger
Institute of Applied Physics, Goethe University, Frankfurt a. M., Germany

## Abstract

Additive manufacturing (or "3D printing") has become a powerful tool for rapid prototyping and manufacturing of complex geometries. As technology is evolving, the quality and accuracy of parts manufactured this way is ever improving. Especially interesting for the world of particle accelerators is the process of 3D printing of stainless steel (and copper) parts. We present the first fully functional IH-type drift tube structure manufactured by metal 3D printing. A 433 MHz prototype cavity has been constructed to act as a proof-of-concept for the technology. The cavity is designed to be UHV capable and includes cooling channels reaching into the stems of the DTL structure. We present the first experimental results for this prototype.

## PROTOTYPE DESIGN AND CONCEPT

The prototype cavity was designed for a resonance frequency of 433.632 MHz, which is a harmonic of the GSI UNILAC operation frequency. In combination with a targeted proton beam energy of 1.4 MeV this scenario allows for a compact accelerator at the limits of feasibility and is therefore a good benchmark for the new approach. The internal drift-tube structure is fully 3D printed from stainless steel (1.4404), see Fig. 1. The cavity frame and lids are manufactured by milling, as those structures are not very complex.

### Beam Dynamics Test Case

The cavity is designed to provide an acceleration of 2.4 MeV, which would equate to an average acceleration gradient of 6.8 MeV/m. A total of six accelerating gaps is modeled after a KONUS zero degree section [1]. The beam dynamics are shown in Fig. 2. A small beam emittance of 0.055 mm mrad and 0.04 keV/u ns is assumed for this test case.

### Mechanical Design

The outer dimensions of the cavity are only 221x206x261 mm (without flanges). A center frame acts as the foundation for the cavity. The center frame provides the precision mount points for the girder-drifttube structures and end-drifttubes. While the end-drifttubes are mounted in vacuum, the girders have a vacuum sealing surface at the bottom. Water channels are included in the girders up to the stems and also in the center frame. Two half shells are mounted on the top and bottom of the center frame. All necessary flanges for vacuum and rf are integrated into the top and bottom half shells. The full construction is shown in Fig. 3.

___
* haehnel@iap.uni-frankfurt.de

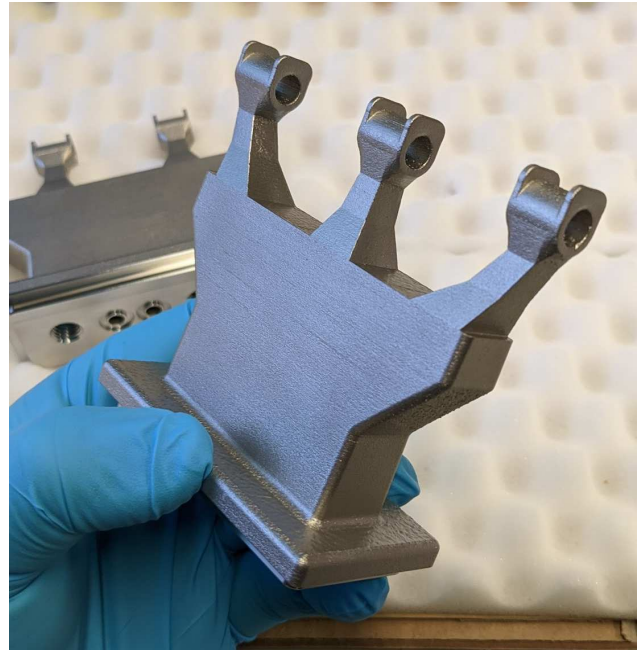

Figure 1: View of the 3D printed girder-drifttube structure including the integrated cooling channels.

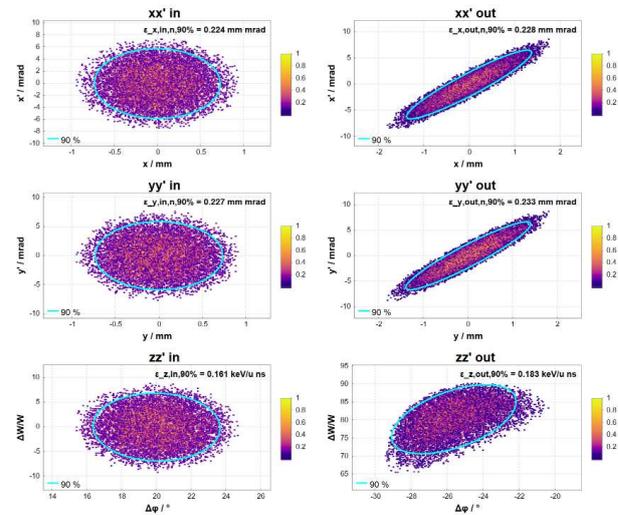

Figure 2: Input and output particle distribution of the prototype cavity beam dynamics.

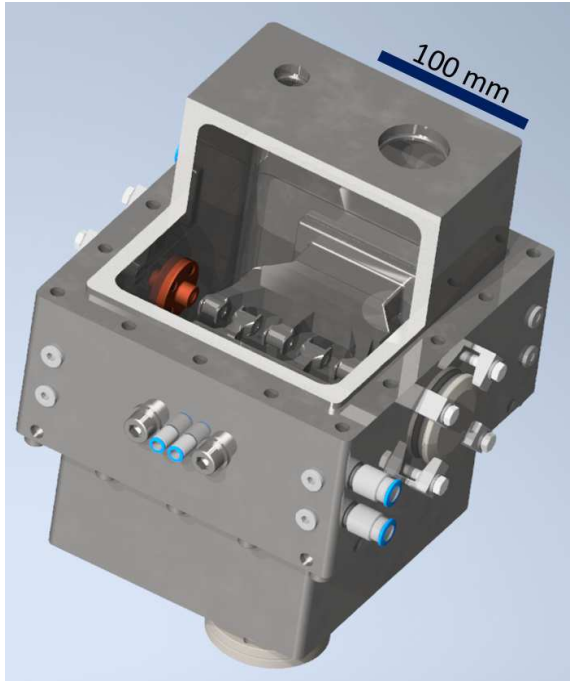

Figure 3: Cut view of the assembled cavity model.

## RF Simulations

The cavity design was optimized for a frequency of 433.632 MHz. The shape of the girder-drifttube structure was also optimized to minimize the need for support structures during the manufacturing process. Figure 4 shows the electric field distribution in the cavity, with the typical characteristics of a βλ/2 IH-type structure.

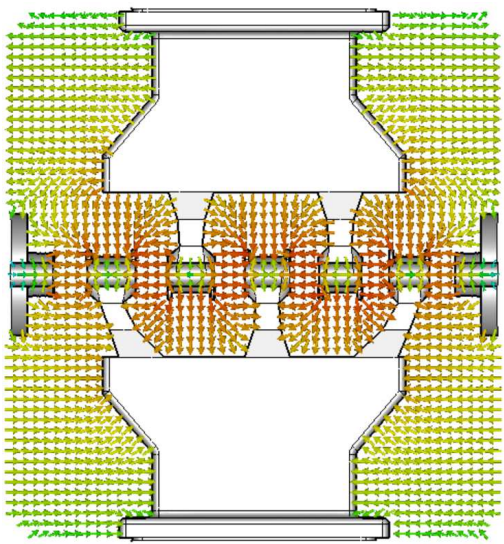

Figure 4: Electric field distribution in the prototype cavity.

## MECHANICAL ACCURACY

Mechanical accuracy of the printed girder-drifttube structure, especially the drifttube geometry, is critical for the field distribution in the cavity and therefore the beam dynamics. The bottom sides of the structures were milled after the printing process to provide a flat sealing surface and o-ring seals for the water channels. In the first step howver, the stems and drifttubes are not machined to evaluate the accuracy provided solely by the printing process. The surface roughness is higher than on machined parts and has to be polished before copper plating.

Table 1: Deviation of Measured Dimensions from the Original Design Values.

| Girder 1 | | Girder 2 | |
| --- | --- | --- | --- |
| $L_{DT,1}$ | +70 µm | $L_{DT,2}$ | +20 µm |
| $L_{gap,1-3}$ | +50 µm | $L_{gap,2-4}$ | +40 µm |
| $L_{DT,3}$ | +70 µm | $L_{DT,4}$ | +90 µm |
| $L_{gap,3-5}$ | +30 µm | | |
| $L_{DT,5}$ | +60 µm | | |

While all the individual drifttube lengths and gap lenghts are within acceptable range ($\Delta L < 100$ µm, see Table 1), all deviations have the same sign and therefore add up. This may be fixed by printing with excess material and shortening the drifttubes after printing using conventional machining. Another possible mitigation would be to simulate the warping of the part during manufacturing and producing a compensating model. This will be studied further, as it would reduce the necessary machining needed after printing.

## WATER FLOW

Both girder-drifttube structures, as well as the center frame of the cavity are watercooled. While the watercooling in the center frame is quite conventional and realized with deep-hole drilled channels, the water cooling in the girders is more complex. To ensure, that the water channels in the girders are working as intended and not blocked by e.g. residual metal powder, some flow measurements were performed. The setup is provided with a water pressure of 8 bar. The tubing used is Festo tubing with an outer diameter of 6 mm. Measurements were performed with a Kobold digital inductive flow meter. The results are summarized in Table 2.

Table 2: Waterflow Measurement

| Scenario | Water Flow |
| --- | --- |
| Source | 10.8 l/min |
| 6mm tubing | 7.2 l/min |
| Girder 1 | 4.8 l/min |
| Girder 2 | 5 l/min |
| Girder 1&2 parallel | 7.6 l/min |

As expected, the complex inner structure of the cooling channels leads to a reduction in flow. However, the measurements show, that significant cooling can be expected for these structures. Future tests with thermal loads will help quantify the cooling capabilities.

## PRELIMINARY VACUUM TESTS

To evaluate the performance of 3D printed components in an UHV environment, preliminary tests were performed with 3D printed beam pipes with KF40 flanges. These pipes were printed in 1.4404 stainless steel and the sealing surfaces on both ends were turned down on a lathe at the IAP workshop to provide a good vacuum seal. The inside of the pipes was left as manufactured. A commercially available conventionally manufactured beam pipe of identical dimensions was acquired from Pfeiffer Vacuum. Both the printed and the conventional pipes were then connected to a small 18 l/s turbomolecular pump and repeatedly pumped and flushed with nitrogen gas. Since these pipes were not cleaned and were used as manufactured, some initial outgassing was expected. The experiment showed some interesting results. For one, as expected with each pumping and flushing cycle, the pump down time to a certain pressure point was taking less time with each iteration. In the end, both pipes performed very similar and the pumping curve as well as the achieved final pressure were indistinguishable within the accuracy limitations of the experiment. Pressure data of the fourth pumping cycle is shown in Fig. 5. The final pressure achieved was about 1 to $2 \cdot 10^{-7}$ mbar for both the conventional and the printed pipe. Similarly promising results were reported in e.g. [2] and [3]. These results certainly show promise for the UHV capability of printed parts. To discern more minute differences between the two samples, a test stand to accurately measure the outgassing rate via the rate-of-rise method is currently being constructed at IAP Frankfurt.

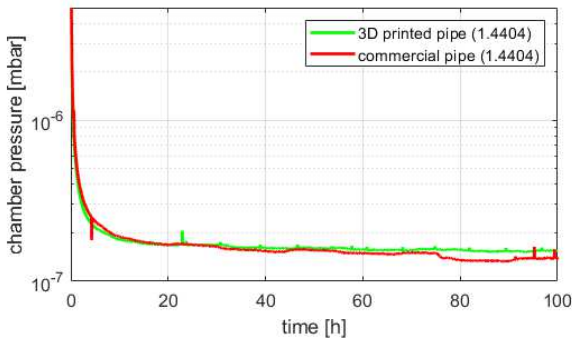

Figure 5: Pumpdown curve of the fourth vacuum cycle of the KF40 test pipes.

## CAVITY VACUUM TEST

The cavity was fully assembled in early May 2021 (see Fig. 6). For vacuum tests, the cavity was attached to a turbomolecular pump (Pfeiffer HiPace80) via one of the top CF40 flanges. A vacuum gauge (Pfeiffer PKR261) was used to measure and log the cavity vacuum. The cavity lids, as well as the girder-drifttube structures were sealed using 1.5 mm aluminum wire.

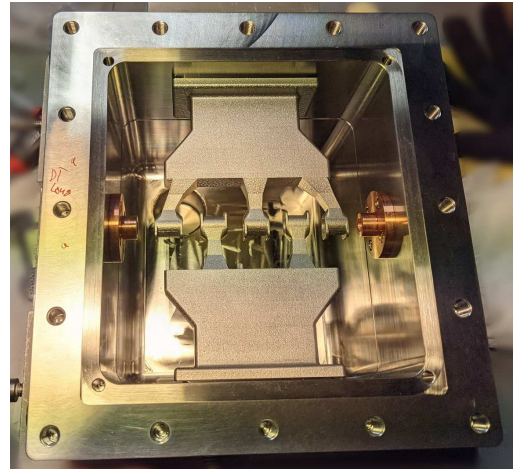

Figure 6: Top view of the cavity during assembly with installed drifttube structure.

At the time of writing, a pressure of $2.97 \cdot 10^{-6}$ mbar was achieved after about 100 hrs of pumping (as shown in Fig. 7). While this result is already promising and the end-pressure is not reached yet, it is belived, that the results can be further improved. Therefore, the cavity will be disassembled once more, the cavity will be cleaned and all metal seals will be redone. All remaining rubber seals on the flanges will also be replaced by copper (CF) or aluminum seals (KF).

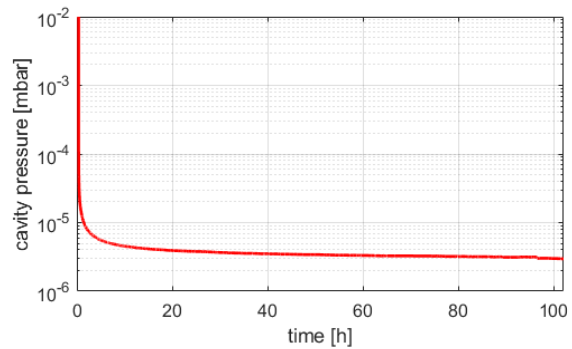

Figure 7: Pumpdown curve of the first vacuum cycle of the fully assembled cavity.

The next steps for the prototype are surface treatment of the girder-drifttube structure and following that copper plating of all inner parts of the cavity. After copper plating, low level rf tests and power rf tests are foreseen.

## CONCLUSION

First vacuum tests show promising results for this proof-of-concept cavity. Mechanical accuracy of the printed parts may be improved by predictive strain simulations or after the fact machining. This prototype is a first step, indicating that stainless steel 3D printed parts can indeed be used as main components for accelerator structures. This opens up many possibilities for more complex designs and possibly previously impossible geometries for particle accelerators.

# REFERENCES

[1] U. Ratzinger, H. Hähnel, R. Tiede, J. Kaiser, A. Almomani, "Combined zero degree structure beam dynamics and applications", *Phys. Rev. Accel. Beams*, vol. 22, no. 11, p 114801, Nov. 2019 `10.1103/PhysRevAccelBeams.22.114801`

[2] G. Sattonnay *et al.*, "Is it Possible to Use Additive Manufacturing for Accelerator UHV Beam Pipes?", in *Proc. IPAC'19*, Melbourne, Australia, May 2019, pp. 2240–2243. `doi:10.18429/JACoW-IPAC2019-WEXXPLS3`

[3] C. R. Wolf, F. B. Beck, L. Franz, and V. M. Neumaier, "3D Printing for High Vacuum Applications", in *Proc. Cyclotrons'19*, Cape Town, South Africa, Sep. 2019, pp. 317–320. `doi:10.18429/JACoW-CYCLOTRONS2019-THC04`